\documentclass[12pt]{iopart}
\usepackage{color}
\usepackage{psfig}
\usepackage{cite}

\begin{document}
\bibliographystyle{iopart-num} 

\title{Anderson Localization of Polar Eigenmodes in Random Planar
  Composites} 

\author{Vadim A. Markel\footnote[1]{E-mail:
  vmarkel@mail.med.upenn.edu}}

\address{Departments of Radiology and Bioengineering, University of
  Pennsylvania, Philadelphia, PA 19104}

\begin{abstract}
  Anderson localization of classical waves in disordered media is a
  fundamental physical phenomenon that has attracted attention in the
  past three decades. More recently, localization of polar excitations
  in nanostructured metal-dielectric films (also known as random
  planar composite) has been subject of intense studies. Potential
  applications of planar composites include local near-field
  microscopy and spectroscopy. A number of previous studies have
  relied on the quasistatic approximation and a direct analogy with
  localization of electrons in disordered solids.  Here I consider the
  localization problem without the quasistatic approximation. I show
  that localization of polar excitations is characterized by algebraic
  rather than by exponential spatial confinement. This result is also
  valid in two and three dimensions. I also show that the previously
  used localization criterion based on the gyration radius of
  eigenmodes is inconsistent with both exponential and algebraic
  localization. An alternative criterion based on the dipole
  participation number is proposed. Numerical demonstration of a
  localization-delocalization transition is given.  Finally, it is
  shown that, contrary to the previous belief, localized modes can be
  effectively coupled to running waves.
\end{abstract}

\date{\today} 
\submitto{JPC}

\maketitle

\section{Introduction}
\label{sec:intro}

Anderson localization (AL) of classical waves in disordered systems is
a fundamental physical phenomenon which takes place in the limit of
strong resonant scattering when the ``photon mean free path'' becomes
the order of or less than one wavelength (the Ioffe-Regel
criterion)~\cite{rossum_99_1}.  At a formal level, AL of
electromagnetic or acoustic waves is similar to localization of
electrons in disordered solids.  There are, however, substantial
physical differences. One such difference is that the motion of
electrons can be finite. In contrast, classical waves can not be, in
principle, indefinitely confined in a finite spatial region. In the
case of electrons, one of the most important physical manifestations
of AL (at zero temperature) is the conductor-insulator
transition~\cite{lee_85_1}. Localization of classical waves is
manifested differently. If we consider an experiment in which
classical waves are transmitted through a disordered slab, an analog
of conductivity is the transmission coefficient. Assuming that the
slab material is non-absorbing, the transmission coefficient can never
turn to zero. However, if the waves are localized in the slab, the
transmitted and reflected fields exhibit spatial variations at
macroscopic scales (much larger than the wavelength) which are
sample-specific (not self-averaging). Emission by localized modes in
random positive-gain media is the basis for operation of random
lasers~\cite{cao_05_1}. Thus, the common feature of localized states
of both electrons and classical waves is that the propagation can not
be described by the Boltzmann transport equation or the diffusion
approximation to the former.

This paper is focused on AL of electromagnetic waves in a random
structure which is distinctly different from either two-dimensional or
three-dimensional random media. Namely, we will consider random planar
composites (RPCs)~\cite{stockman_01_1}. The RPCs are made of small
three-dimensional scatterers which are randomly distributed in a thin
planar layer.  Thus, the electromagnetic interaction in this system is
essentially three-dimensional while the geometry is two-dimensional.
The RPCs have attracted considerable recent attention due to the
variety of potential applications, including surface-enhanced Raman
spectroscopy of
proteins~\cite{drachev_04_1,drachev_05_1,drachev_05_2}. The physical
implication of AL of electromagnetic waves in RPCs can be best
understood by considering an experiment in which the sample is excited
by a near-field probe. If the electromagnetic states in the sample are
strongly localized (at the particular electromagnetic frequency), the
surface plasmon induced by the tip will also be localized and not
spread over the entire sample. It must be emphasized, however, that
there are other mechanisms that can lead to spatial exponential decay
of surface plasmons. This includes decay due to absorption (Ohmic
losses in the material and

The localization-delocalization transition is expected to play a
crucial role in the near-field tomographic imaging techniques of
Refs.~\cite{carney_01_2,carney_04_1}. If the states are localized,
each near-field measurements will be sensitive only to the local
environment of the the tip, while in the opposite case, it would be
sensitive to the structure of the sample far away from the tip.  The
relation between AL and transport of surface plasmons is discussed in
detail in Section~\ref{sec:concept}. 

The possibility and nature of AL of electromagnetic excitations in the
RPCs have been investigated theoretically and
numerically~\cite{stockman_01_1,genov_05_1}. While the conclusions
given in these two references are somewhat conflicting, the respective
theoretical approaches share some common features. Most importantly,
localization of the SP eigenmodes was studied in the quasistatic
approximation. However, AL is, essentially, an interference
phenomenon~\cite{rossum_99_1}. Therefore, account of retardation is
essential for its proper understanding.  Second, a definition of
localization length of a mode based on its ``radius of gyration'' was
adopted in Refs.~\cite{stockman_01_1,genov_05_1}.  Here I argue that
this definition, as well as the one based on radiative quality factor
(mode lifetime)~\cite{rusek_97_1,rusek_00_2} can not be applied to the
electromagnetic localization problem in the RPCs.

Below, I discuss a number of important points concerning AL of
classical waves, some of which are applicable only to the RPCs and
some are more general. I also provide a numerical demonstration of the
Anderson transition in the RPCs. To this end, I use a simple but
physically relevant model of small spherical inclusions of diameter
$D$ embedded in a transparent dielectric host medium and randomly
distributed in a plane inside an $L\times L$ box.  Essentially, this
is the model used in Refs.~\cite{rusek_97_1,rusek_00_2}.
However, I work in a different physical regime and use a different
definition for localization.  Most of the numerical examples shown
below were obtained in the limit $D\ll \ell \ll \lambda \ll L$, where
$\ell$ is the average inter-particle distance and $\lambda$ is the
wavelength.

\section{Physical Model}
\label{sec:model}

Consider an infinite transparent dielectric host with $N$ small
identical spherical inclusions of diameter $D$ randomly distributed in
an $L \times L$ box in the $xy$-plane. The inclusions interact with a
plane linearly-polarized electromagnetic wave with the wave number
$k=n\omega/c=2\pi/\lambda$, where $n$ is the host index of refraction,
$\omega$ is the electromagnetic frequency. We assume that $\lambda\gg
D$, but the relation between $\lambda$ and $L$ is arbitrary. Thus, we
do not use the {\em quasistatic approximation}. However, we do use the
{\em dipole approximation} which is accurate in the limit of small
density of inclusions and $D \ll \lambda$. Assuming that the
inclusions are non-magnetic, the electric dipole moments ${\bf d}_i$
($i=1,\ldots,N$) induced in each spherical inclusion satisfy the
coupled-dipole equation

\begin{equation}
\label{CDE}
{\bf d}_i =\alpha\left[{\bf E}_{\rm inc}\exp(i{\bf k}_{\rm inc}
  \cdot{\bf r}_i) + \sum_{j\neq i} G({\bf r}_i,{\bf r}_j) {\bf d}_j
\right] \ , 
\end{equation}

\noindent
where ${\bf E}_{\rm inc}$ is the amplitude of the incident wave,
$\vert {\bf k}_{\rm inc} \vert = k$, $\alpha$ is the polarizability of
inclusion, ${\bf r}_i$ is the radius-vector of the $i$-th inclusion
and $G({\bf r}_i,{\bf r}_j)$ is the dyadic Green's function for the
electric field in a homogeneous infinite host medium given by

\begin{eqnarray}
\label{G_tens}
&&G({\bf r}_i,{\bf r}_j) = k^3 \left[ A(kr_{ij})I + B(kr_{ij})
\hat{\bf r}_{ij}\otimes\hat{\bf r}_{ij} \right] \ ,
\\
\label{A_x}
&&A(x) = [x^{-1} + ix^{-2} - x^{-3}]\exp(ix) \ ,
\\
\label{B_x}
&&B(x) = [-x^{-1} - 3ix^{-2} + 3x^{-3}]\exp(ix) \ .
\end{eqnarray}

\noindent
Here $I$ is the unit dyadic, ${\bf r}_{ij}={\bf r}_j - {\bf r}_i$,
$r_{ij}=\vert {\bf r}_{ij} \vert$, $\hat{\bf r}_{ij} = {\bf
  r}_{ij}/r_{ij}$ and $\otimes$ denotes tensor product.  

The system of equation (\ref{CDE}) can be written in operator form as

\begin{equation}
\label{CDE_oper}
\vert d \rangle = \alpha(\vert E_{\rm inc} \rangle + W \vert d \rangle
) \ .
\end{equation}

\noindent
Here the Cartesian components of all dipole moments are given by
$d_{i,\sigma}=\langle i, \sigma \vert d \rangle$, where $i=1,\ldots ,
N$ and $\sigma=x,y,z$. The above relation defines the orthonormal
basis $\vert i,\sigma \rangle$. 

The $3N$-dimensional matrix $W$ is complex symmetric and, hence,
non-Hermitian. Since such matrices are not very common in physics, a
brief review of their spectral properties is adduced. Eigenvalues of
complex symmetric matrices are, generally, complex. The eigenvectors
form a complete (but not orthonormal) basis unless the matrix is {\em
  defective}. A matrix is defective if one of its eigenvectors is {\em
  quasi-null}, e.g, its dot product with itself (without complex
conjugation) is zero (see below). The geometric multiplicity of a
defective matrix is less than its algebraic multiplicity.
Non-degenerate symmetric matrices are all non-defective. A matrix can
be defective as a result of {\em random degeneracy}. The probability
of such event is, however, vanishingly small. Below, we assume that
$W$ is non-defective, which was the case in all numerical simulations
shown below. Further, let $\vert a \rangle$ and $\vert b \rangle$ be
two distinct eigenvectors of $W$ with components $\langle n \vert a
\rangle = a_n$ and $\langle n \vert b \rangle = b_n$. The usual
orthogonality condition $\langle a \vert b \rangle = 0$ is replaced by

\begin{equation}
\label{a_b_orthogonality}
\sum_n a_n b_n \equiv \langle \bar{a} \vert b \rangle = 0 \ .
\end{equation}

\noindent
Note that the bilinear form in the above formula is defined without
complex conjugation. Such forms are called quasi-scalar products and
are denoted by $\langle \bar{a} \vert b \rangle$, in contrast to the
true scalar product $\langle a \vert b \rangle$. The quasi-scalar
product of a vector with itself, $\langle \bar{a} \vert a \rangle$ is,
generally, a complex number, possibly zero. A vector whose
quasi-scalar product with itself is zero is called quasi-null. At the
same time, each eigenvector (including quasi-null vectors) can be
normalized in the usual way, so that $\langle a \vert a \rangle = 1$.

Let $w_n$ and $\vert \psi_n \rangle$ be the set of $3N$ eigenvalues
and eigenvectors of $W$. We assume here that $\vert \psi_n \rangle$
are normalized so that $\langle \psi_n \vert \psi_n \rangle = 1$.
However, the quasi-scalar product $\langle \psi_n \vert \psi_n
\rangle$ is, in general, a complex number. We can use the
orthogonality rule (\ref{a_b_orthogonality}) to obtain the spectral
solution to (\ref{CDE_oper}):

\begin{equation}
\label{d_spectral}
\vert d \rangle = \sum_n {{\vert \psi_n \rangle \langle
  \bar{\psi_n} \vert E_{\rm inc} \rangle} \over {\langle \bar{\psi_n} \vert
  \psi_n \rangle (z - w_n)}} \ ,
\end{equation}

\noindent
where $z=1/\alpha$. We note that this spectral solution has been
obtained assuming there are no quasi-null eigenvectors. In the
opposite case, spectral solution can not be obtained.

For non-absorbing inclusions, ${\rm Im}z=-2k^3/3$. This equality can
be readily obtained by observing that the extinction cross section
$\sigma_e = 4\pi k {\rm Im}\alpha$ and the scattering cross section
$\sigma_s = (8\pi k^4/3)\vert \alpha \vert^2$ must be equal in the
absence of absorption~\cite{markel_92_1} ($\alpha = 1/z$).
Analogously, in the case of finite absorption, we have ${\rm Im}z <
-2k^3/3$.  Consequently, energy conservation
requires~\cite{markel_95_1} that ${\rm Im}w_n \geq -2k^3/3$. The
eigenstates with ${\rm Im}w_n = -2k^3/3$ are
non-radiating~\footnote{Because, in this case, and assuming the
  particles are non-absorbing, the terms ${\rm Im}z$ and ${\rm Im}w_n$
  in the denominator of (\ref{d_spectral}) cancel each other.
  Physically, this corresponds to cancellation of radiative reaction
  due to the interference effects.  See
  Refs.~\cite{markel_92_1,markel_95_1} for more details.}  while the
eigenstates with ${\rm Im}w_n \sim N(2k^3/3)>0$ are super-radiating.
The radiative quality factor of the mode is defined as
$Q_n=1/\gamma_n$, $\gamma_n={\rm Im}[w_n/(2k^3/3)] + 1$.  For a
non-radiating state, $\gamma_n\rightarrow 0$ and $Q_n \rightarrow
\infty$. The {\em coupling constant} $f_n$ for the $n$-th mode is
defined as $f_n=\langle E_{\rm inc} \vert \psi_n \rangle \langle
\bar{\psi}_n \vert E_{\rm inc} \rangle [\vert {\bf E}_{\rm inc}\vert^2
\langle \bar{\psi}_n \vert \psi_n \rangle]^{-1}$; the $f_n$'s satisfy
the sum rule $\sum_n f_n = N$.

\section{The Concept of Anderson Localized for Polar Eigenmodes}
\label{sec:concept}

In this Section I discuss in more detail the concept of AL of polar
eigenmodes and the relation between localization and transport
properties. In particular, a rationale is given for studying spatial
properties of the eigenmodes which depend only on the sample geometry
but not on the material properties of the host medium or inclusions.

It is well known that the electromagnetic problem of two-component
composites, if solved within the quasistatics, allows for an effective
separation of material properties of the constituents and the geometry
of the composite. This idea goes back to the Bergman-Milton spectral
theory of composites~\cite{bergman_pr} and has been used in many
different settings. For example, dipolar (more generally, multipolar)
excitations in aggregated spheres were studied in the 1980-ies by
Fuchs, Claro and co-authors~\cite{rojas_86_1,fuchs_89_1,claro_91_1}
using the spectral approach analogous to the Bergman-Milton spectral
theory of composites.

I have extended the quasistatic spectral theory of
Refs.~\cite{bergman_pr,rojas_86_1,fuchs_89_1,claro_91_1} to the case
of samples which are not small compared to the wavelength, e.g., when
the effects of retardation are important in
Refs.~\cite{markel_95_1,markel_97_1}. In this case, similarly to the
quasistatics, electromagnetic eigenstates of a two-component mixture
or composite can be defined. These eigenstates turn out to be
independent of the material properties of the constituents, but,
unlike in the quasistatic limit, depend explicitly on $k=\omega/c$.
Yet, at a fixed electromagnetic frequency $\omega$, the spatial
properties of the eigenstates can be studied irrespectively of the
material properties of the constituents.  In particular, one can argue
that in extended systems the eigenstates can be either localized on
several inclusions or delocalized over the whole sample, completely
independently of the material properties. This was the point of view
taken in Refs.~\cite{stockman_01_1,genov_05_1}.

The connection between propagation of surface plasmon excitations in
the system and the localization properties of the eigenstates can be
established by examining the spectral solution (\ref{d_spectral}) for
the case of local excitation (e.g., by a near-field tip) at the site
${\bf r}_1$. If the point of observation is located at the site ${\bf
  r}_2$ (e.g., another near-field tip operating in the collection
regime), the amplitude of the measured signal is proportional to the
following Green's function:

\begin{equation}
\label{G_def}
{\mathcal D}({\bf r}_2,{\bf r}_1) = \sum_n \frac{f_n({\bf r}_2,{\bf r}_1)}{z -
  w_n} \ ,
\end{equation}

\noindent
where

\begin{equation}
\label{numerator}
f_n({\bf r}_2,{\bf r}_1) = \frac{\langle {\bf r}_2 \vert \psi_n \rangle \langle \bar{\psi}_n \vert {\bf
  r}_1 \rangle}{\langle \bar{\psi}_n \vert \psi_n \rangle }  \ .
\end{equation}

\noindent
Let us assume that at a given electromagnetic frequency $\omega$ there
are resonance modes, e.g., such mode that ${\rm
  Re}[z(\omega)-w_n(\omega)]\approx 0$. This is, obviously, not always
the case, and the above condition depends, in particular, on the
material properties of the medium. However, if the resonance
excitation of the system is, in principle, possible, summation in
(\ref{G_def}) can be restricted to resonant modes. In this case,
propagation of surface plasmons in the system is governed by the
spatial dependence of the functions $f_n({\bf r}_2,{\bf r}_1)$, where
$n$ indexes only resonant modes.

In a strictly periodic infinite system, all eigenmodes are delocalized
plane waves, which follows from general symmetry considerations. These
delocalized modes form a truly {\em continuous} spectrum and can be
labeled by a continuous index $q$, where $q$ is the wave vector in the
first Brillouin zone of the lattice (here the dimension of $q$ is not
specified). Note that the eigenmodes that belong to the continuous
spectrum are non-normalizable in the usual sense. This means that the
scalar product $\langle \psi_q \vert \psi_q \rangle$ does not exists.
Instead, the usual delta-function normalization $\langle \psi_q \vert
\psi_{q^{\prime}} \rangle =\delta(q-q^{\prime})$ must be used.
Obviously, functions $f_q({\bf r}_2,{\bf r}_1)$ that correspond to the
delocalized eigenstates are non-decaying as the distance between the
points ${\bf r}_1$ and ${\bf r}_2$ increases.

At this point, two important comments must be made. First, the fact
that the the eigenmodes $\vert \psi_q \rangle$ are delocalized plane
waves and the corresponding functions $f_q({\bf r}_2,{\bf r}_1)$ are
non-decaying does not necessarily imply that there can be no
exponential spatial decay of surface plasmon excitations in the
system. This will be illustrated later in this Section with several
examples. Second, eigenvalues in a strictly periodic and {\em
  non-chiral} system are invariant with respect to the change
$q\rightarrow -q$ and, therefore, doubly degenerate. This does not
lead to defectiveness of the operator $W$ but the degenerate modes
must be appropriately orthogonalized.

If we now introduce disorder and break the translational symmetry of
an infinite sample, the eigenmodes $\vert \psi_n \rangle$ are no
longer plane waves. Still, the modes can be either localized or
delocalized. The fundamental property of localized modes is that such
modes are square integrable in the sense that $\langle \psi_n \vert
\psi_n \rangle < \infty$ (in fact, localized modes can always be
normalized so that $\langle \psi_n \vert \psi_n \rangle =1$) and,
therefore, belong to a {\rm discrete} spectrum. There are two
consequences of this fact. First, a localized mode can be
characterized by a ``center of mass'' or a point near which it is
localized, which we denote by ${\bf R}$. Second, localized modes can be
labeled by a countable index. It is natural to use ${\bf R}$ itself
as a label. Let all resonant modes be localized. Then the Green's
function (\ref{G_def}) can be written as

\begin{equation}
\label{G_loc}
{\mathcal D}({\bf r}_2,{\bf r}_1) = \sum_{{\bf R}} \frac{f_{{\bf R}}({\bf
    r}_2,{\bf r}_1)}{z - w_n} \ , 
\end{equation}

\noindent
where summation is extended over resonant modes localized near the
point ${\bf R}$. It is clear from inspection of Eq.~(\ref{G_loc})
and the definition (\ref{numerator}) that the above Green's function
is spatially decaying when $\vert {\bf r}_1 - {\bf r}_2\vert
\rightarrow \infty$, and that the rate of this decay depends on the
spatial decay of the eigenmodes.

However, it can be readily seen that the spatial decay of localized
eigenmodes does not need to be exponential. This is because the
essential requirement of localization $\langle \psi_n \vert \psi_n
\rangle < \infty$ can be satisfied even if the decay is algebraic.
More specifically, it is sufficient that $\psi_{\bf R}({\bf r}) =
\langle {\bf r} \vert \psi_{{\bf R}} \rangle$ decays faster than
$1/\vert {\bf r} - {\bf R} \vert^{d/2}$, where $d$ is the
dimensionality of the sample.  Accordingly, the spatial decay of the
Green's function (\ref{G_loc}) can be slower than exponential. In
fact, I argue in this paper that electromagnetic eigenstates can not
be localized exponentially. This is a consequence of algebraic decay
of the free-space Green's function (\ref{G_tens}) in a transparent
host medium. More specifically, if $\vert \psi_n \rangle$ is an
exponentially localized eigenstate, the equation $W\vert \psi_n
\rangle = w_n \vert \psi_n \rangle$ can not be satisfied due to the
fact that $\langle {\bf r} \vert W \vert \psi_n \rangle$ is
asymptotically algebraic and $\langle {\bf r} \vert \psi_n \rangle$ is
asymptotically exponential. The impossibility of exponential
localization of electromagnetic eigenmodes can be also viewed as a
consequence of the fact that there are no bound states for light.

The above arguments apply to infinite samples.  However, any numerical
computation is restricted to finite systems.  Modes of a finite system
are always discrete and have a finite $L_2$ norm. The difference
between the modes which remain discrete in the thermodynamic limit and
the modes that become delocalized is then revealed by examining the
rate at which these modes decay. The amplitudes of localized modes
decay (away from their ``center of mass'' ${\bf R}$) faster than
$1/\vert {\bf r} - {\bf R}\vert^{d/2}$. Consequently, a localized
mode centered at a point ${\bf R}$ which is sufficiently far from
the sample boundaries is insensitive to any change in the sample
overall size or shape. In contrast, delocalized modes always remain
sensitive to the exact shape of the boundaries. For example, a
delocalized mode would change substantially if the system overall size
is doubled while a localized mode will be insensitive to such change,
as along as it is centered sufficiently far from the original
boundaries.

In light of the above, it is interesting to examine the applicability
of a localization criterion which based on the eigenmode gyration
radius, $\xi_n$ (defined in Section~\ref{sec:loc_states} below). This
parameter was used in Refs.~\cite{stockman_01_1,genov_05_1} within the
quasistatic approximation. An eigenmode was considered to be localized
if $\xi_n \ll L$, where $L$ is the sample characteristic size. The
above inequality was claimed to be a consequence of exponential
localization. First, we note that the condition $\xi_n \ll L$ is not
equivalent to (actually, is weaker than) exponential localization of
the eigenmode. As is shown in Section~\ref{sec:loc_states} below, it
is sufficient that the eigenstate exhibits spatial decay faster than
$1/r^{d/2+1}$ for the above inequality to hold. But more importantly,
application of this criterion can lead to a grossly inaccurate
conclusion about delocalization of an eigenstate when it is, in fact,
strongly localized. The reason for this is numerical. Indeed, even
though localized eigenstates belong to a countable discrete spectrum,
the spacings between eigenvalues $w_n$ can be arbitrarily small and
tend to decrease with the sample size. Thus, two localized eigenstates
with centered at different points ${\bf R}_1$ and ${\bf R}_2$ can have
very close eigenvalues $w_1 \approx w_2$. Numerically, these two
eigenstates are {\em quasi-degenerate}. That means that any linear
superposition of these two eigenstates is also an eigenstate (within
numerical precision of the computer). Namely, if $\vert \psi_{{\bf
    R}_1} \rangle$ and $\vert\psi_{{\bf R}_2} \rangle$ are two
quasi-degenerate eigenstates, then the linear combinations

\begin{eqnarray}
\vert \psi_{\pm} \rangle = \frac{1}{\sqrt{2}}(\vert\psi_{{\bf
    R}_1} \rangle \pm \vert\psi_{{\bf R}_2} \rangle ) \nonumber
\end{eqnarray}

\noindent
are also eigenstates (within the numerical precision). But the
gyration radii of $\vert\psi_{{\bf R}_1} \rangle$, $\vert\psi_{{\bf
    R}_2} \rangle$, on one hand, and of $\vert\psi_{{\bf r}_{\pm}}
\rangle$, on the other, might be very different. Thus, if the first
two states are localized with gyration radii $\xi_1,\xi_2\ll L$, the
gyration radii of the eigenmodes $\vert\psi_{{\bf r}_{\pm}} \rangle$
can be of the order of $\vert {\bf R}_1 - {\bf R}_2 \vert \sim L$.  I
argue below that a more reliable numerical criterion of localization
can be based on the so-called {\em participation number}.

Thus, we have established that geometrical properties of
electromagnetic eigenmodes play a fundamental role in AL. These
eigenmodes are completely defined by the sample geometry and are
independent of the material properties, as long as there is no more
than two constituents in the medium. Therefore, following
Refs.~\cite{stockman_01_1,genov_05_1} we focus on the properties of
the eigenmodes and do not consider a specific material- and
frequency-dependent model for the spectral parameter $z$. It should be
noted, however, that polarization or electric field can be
exponentially confined in space due to several reasons other than AL.
The physical phenomena that lead to such confinement can strongly
depend on material properties of the medium and should not be confused
with AL. In the remainder of this Section, we briefly review several
such phenomena.

The first relevant example is exponential decay of evanescent waves.
Evanescent decay can lead, for example, to exponential confinement of
waves in one-dimensional periodic layered media (photonic crystals).
However, evanescent waves are exponentially localized only in one
selected direction and are oscillatory in any direction orthogonal to
the former.

The second example is decay due to dissipation. It is well known that
a superposition of perfectly delocalized states can result in an
exponentially decaying wave. Consider a superposition of
one-dimensional waves with continuous wave-numbers $q$ (e.g., in a
one-dimensional periodic system):

\begin{equation}
\label{E_q}
d(x) = \int_{-\pi/h}^{\pi/h} \frac{h dq}{2\pi} \frac{e^{iqx}}{z - w(q)}
\end{equation}

\noindent
This equation would describe, for example, propagation of a surface
plasmon along a linear periodic chain of nanospheres of period $h$ and
integration is extended over the first Brillouin zone of the
lattice~\cite{markel_submitted}. It can be shown
exactly~\cite{burin_04_1} that for $q>k=\omega/c$, ${\rm
  Im}w(q)=-2k^3/3$. (This is known as the light-cone condition; SP is
not coupled to running waves because momentum of the photon can not be
conserved. Due to the same reason, a propagating SP with $q>k$ does
not radiate.)  We now make the usual quasi-particle pole
approximation, namely

\begin{equation}
\label{QPP_approx}
w(q) \approx {\rm
  Re} w(q_0) + (q-q_0) \left\vert \frac{\partial {\rm Re} w(q)}{\partial q}
\right \vert_{q=q_0} - i\frac{2k^3}{3} \ , 
\end{equation}

\noindent
where $q_0$ is, by definition, the solution to ${\rm Re}[z-w(q_0)]=0$,
and find that the dipole moment decays in space as $exp(-\vert x
\vert/\ell)$, where the exponential scale is

\begin{equation}
\label{ell_def}
\ell = \frac{1}{\delta} \left \vert \frac{\partial {\rm
      Re}w(q)}{\partial q} \right \vert_{q=q_0} \ .
\end{equation}

\noindent
and

\begin{equation}
\label{delta_def}
\delta = -[{\rm Im}z + 2k^3/3]
\end{equation}

\noindent
is a non-negative parameter characterizing the absorption strength of
one isolated nanosphere. In general, it can be shown that $\delta=0$
in non-absorbing particles (whose dielectric function
$\epsilon(\omega)$ is purely real at a given frequency $\omega$). The
important point is that we have obtained exponential decay in space,
even though we have superimposed delocalized states (running waves
$\exp(iqx)$). And notice that the localization length in this case is
directly defined by material properties through the relaxation
constant $\delta$. Numerical verification of Eq.~\ref{ell_def} is
given in Ref.~\cite{markel_submitted}.

A third type of localization happens in the off-resonant or weak
interaction limit. Mathematically, this takes place when $z$ is large
and we can neglect the term $w(q)$ in the denominator of (\ref{E_q})
(or replace $z-w(q)$ by a $q$-independent constant). This will result
in polarization localized as ${\rm sinc}(x/h)$, or in the discrete
case, as the Kronecker delta-symbol.  Physically, this is
manifestation of the fact that surface plasmons can not propagate when
the distance between polarizable particles is too large, or the
electromagnetic frequency is very far from the resonance.

\section{Numerical Methods}
\label{sec:numeric}

All numerical results shown below were obtained by direct numerical
diagonalization of the interaction operator $W$. A FORTRAN code has
been written to model a random RPC and to compute elements of the
matrix $W$ as well as its eigenvectors and eigenvalues. RPCs were
generating by randomly placing particles inside a two-dimensional
$L\times L$ box with the only requirement that each particle does not
approach any of the previously placed particles closer that one
diameter $D$. An attempt which did not satisfy this requirement was
rejected and the process was repeated until the total of $N$ particles
were placed inside the box. The box was considered to be embedded in
infinite space; thus no periodic or other boundary conditions
different from the usual scattering conditions at infinity were
applied.

Diagonalization (computation of eigenvectors and eigenvalues of $W$)
was accomplished by utilization of the LAPACK subroutine ZGEEV.
Recall that $W$ is a $3N \times 3N$ complex symmetric matrix. In the
case of RPC, it is also a block matrix. It contains an $N\times N$
block whose eigenvectors correspond to excitations polarized
perpendicular to the RPC and an $2N \times 2N$ block whose
eigenvectors correspond to in-plane excitations. Each block can be
diagonalized independently.  The code was compiled and executed on an
HP rx4640 server (1.6GHz Itanium-II cpu) with the Intel's FORTRAN
compiler and MKL mathematical library.  Diagonalization time (for
serial execution) for a matrix of the size $M$ scaled approximately as
$30 \times (M/1000)^3{\rm sec}$. The relatively large computational
time is a consequence of $W$ not being Hermitian. Diagonalization of
Hermitian matrices is much more computationally efficient. It should
be noted that the procedure always returned $M$ linearly-independent
eigenvectors. There were no quasi-null eigenvectors and,
correspondingly, $W$ was not defective.

\section{Properties of Localized States}
\label{sec:loc_states}

We first discuss how to determine if a certain eigenstate $\vert\psi_n
\rangle$ is localized in the Andersen sense. Various definitions of
localizations that has been used for electrons in disordered solids
are reviewed in Ref.~\cite{kramer_93_1}. In the literature on
localization of polar (electromagnetic) modes in disordered
composites, two approaches have been adopted. The first approach is
based on the assumption that the localization length $\xi_n$ is the
order of the gyration radius of the mode, $\langle r^2 \rangle_n -
\langle {\bf r} \rangle^2_n$, where $\langle \ldots \rangle_n$ denotes
a {\em weighted} average~\cite{stockman_01_1,genov_05_1}. More
specifically, we define for the $n$-th mode and for each inclusion
located at ${\bf r}={\bf r}_i$ the weight

\begin{equation}
\label{m_n_def}
m_n({\bf r}_i) = \sum_{\sigma}\langle \psi_n \vert i,\sigma \rangle
\langle i,\sigma \vert \psi_n \rangle \ , 
\end{equation}

\noindent
where $\sigma=x,y,z$ labels the Cartesian components of
three-dimensional vectors. Note that the argument ${\bf r}_i$ in the
expression $m_n({\bf r}_i)$ is discrete. The basis $\vert i, \sigma
\rangle$ was defined after Eq.~(\ref{CDE_oper}). Since $\vert i,
\sigma \rangle$ form an orthonormal basis, we have $\sum_{i,\sigma}
\vert i,\sigma \rangle \langle i, \sigma \vert = 1$. We then recall
that the eigenvectors $\vert \psi_n \rangle$ are normalized so that
$\langle \psi_n \vert \psi_n \rangle = 1$ and find that the weights
satisfy the following sum rule:

\begin{equation}
\label{m_n_norm}
\sum_i m_n({\bf r}_i) = 1 \ .
\end{equation}

\noindent
However, note that, in general, $\sum_n m_n({\bf r}_i) \neq 1$, with
the equality holding only in the quasistatic limit. This is because
the basis of eigenvectors $\vert \psi_n \rangle$ is not orthonormal
(although it is complete). Then, according to
Refs.~\cite{stockman_01_1,genov_05_1}, the localization length $\xi_n$
for the $n$-th eigenmode is defined as

\begin{equation}
\label{xi_def}
\xi_n^2 = \sum_i m_n({\bf r}_i)r_i^2 - \left( \sum_i m_n({\bf r}_i)
  {\bf r}_i \right)^2 \ .
\end{equation}

\noindent
This definition is implicitly based on the assumption of exponential
localization. However, exponential decay (in space) of the weights
$m_n({\bf r}_i)$ is impossible for classical waves. This follows
already from the fact that the unperturbed Green's function
(\ref{G_tens}) in non-absorbing transparent host media decays
algebraically rather than exponentially. However, the exponential
localization is not an absolute requirement for AL.  Indeed, the
essential feature of strongly localized states is that such states are
discrete and can be, therefore, labeled by countable
indices~\cite{anderson_78_1}.  These indices can be associated, for
example, with localization regions which are, by definition,
countable. As a consequence, the localized states are normalized in
the usual sense, implying that $\int m_n(r)r^{d-1}dr$ converges at the
upper limit, where $d$ is the dimensionality of embedding space and we
have approximated summation over discrete variables ${\bf r}_i$ by
integration over the volume of the sample. Note that for the RPCs
$d=2$, even though the interaction is three-dimensional. Therefore, a
state is localized if the weights decay faster than $1/r^d$. In
contrast, delocalized states belong to the true continuum (in an
infinite system). Such states can not be normalized in the usual sense
but instead satisfy $\langle \psi_\mu \vert \psi_\nu \rangle =
\delta(\mu -\nu)$, where $\mu$ and $\nu$ are continuous variables.
Consequently, the above integral is diverging for delocalized states.
Now consider the localization length defined by (\ref{xi_def}).
Without loss of generality, we can assume that the center of mass (the
second term in (\ref{xi_def})) is zero. The first term converges if
$m_n(r)$ decays faster than $1/r^{d+2}$ and diverges otherwise. Thus,
the requirement that $\xi_n \ll L$ is, in fact, much stronger than is
necessary for localization (yet, is still weaker than the requirement
of exponential localization). That is, some modes which are truly
localized in the Anderson sense will appear to be delocalized
according to the definition (\ref{xi_def}).

To illustrate this point, I introduce a different localization
parameter. Let 

\begin{equation}
\label{M_n_def}
M_n = \left[ \sum_{i=1}^N m_n^2({\bf r}_i) \right]^{-1} \ .
\end{equation}

\noindent
We will refer to $M_n$ as the {\em participation number} of the $n$-th
eigenmode. It is directly analogous to the participation number
defined as the inverse second moment of the probability density or the
inverse fourth moment of the wave function~\cite{kramer_93_1}. It can
be seen that, given the constraint (\ref{m_n_norm}), possible values
of $M_n$ lie in the interval $1 \leq M_n \leq N$. Thus, for example,
if all the weights are equal, $m_n({\bf r}_i)=1/N$, we have $M_n=N$. If
the mode is localized on just one inclusion $i=i_0$, so that $m_n({\bf
  r}_i)=\delta_{i,i_0}$, we have $M_n=1$.  In general, a mode can be
considered localized if $M_n \ll N$.

\begin{figure}
\centerline{\psfig{file=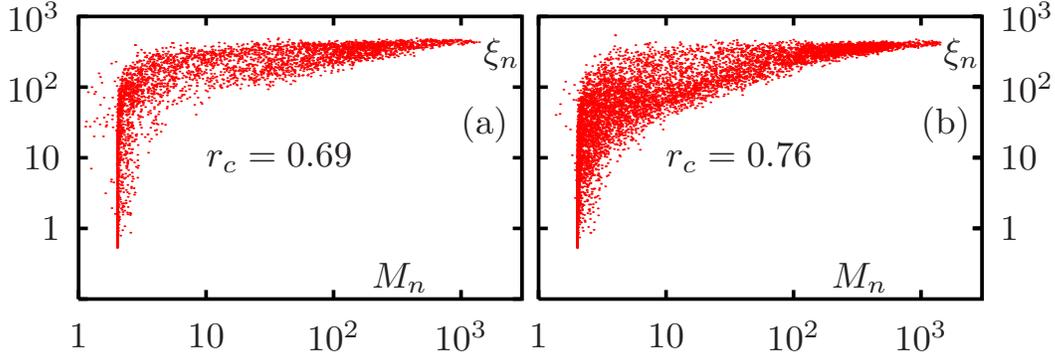,width=14.0cm,bbllx=165bp,bblly=620bp,bburx=450bp,bbury=742bp,clip=t}}
\caption{Gyration radii $\xi_n$ vs the the participation numbers
  $M_n$ of all modes with polarization perpendicular to the plane (a)
  and parallel to the plane (b). Parameters: $N=4000$, $L=1000D$ and
  $\lambda=100D$. Numerical value of the correlation coefficient $r_c$
  between $M_n$ and $\xi_n$, calculated for all modes, is indicated in
  each plot.}
\label{fig:1}
\end{figure}

In Fig.~\ref{fig:1}, the participation number $M_n$ is compared to
$\xi_n$ for all modes in an RPC consisting of $N=4,000$ inclusions.
While there is positive correlation between $M_n$ and $\xi_n$ (the
correlation coefficient $r_c$ is indicated in the figure), it can be
readily seen that many modes which are localized in the sense that
$M_n \ll N$ have the gyration radius of the order of $L$.  Thus, while
some correlation between $M_n$ and $\xi_n$ exists, there is virtually
no such correlation when $M_n \ll N$.

\begin{figure}
\centerline{\psfig{file=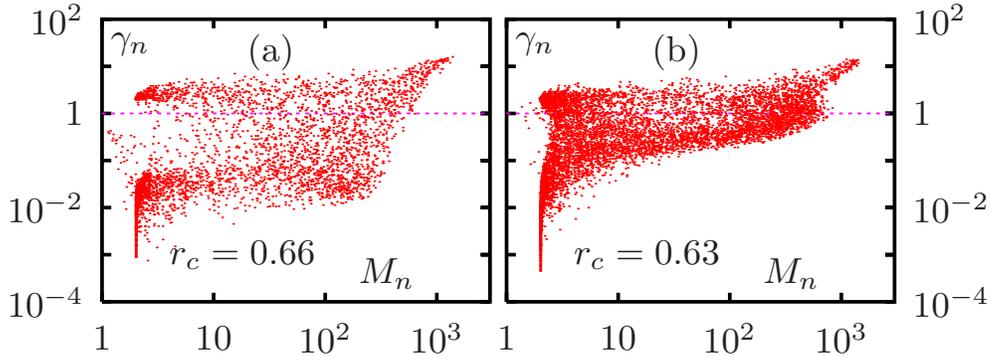,width=14cm,bbllx=165bp,bblly=620bp,bburx=450bp,bbury=742bp,clip=t}}
\caption{Inverse radiative quality factors $\gamma_n$ vs $M_n$. Same
  parameters as in Fig.~\ref{fig:1}. Numerical value of the correlation
  coefficient $r_c$ between $M_n$ and $\gamma_n$ is indicated in each
  plot.}
\label{fig:2}
\end{figure}

The second approach to defining localization which has been used in
the literature is based on the eigenmode radiative quality
factor~\cite{rusek_97_1,rusek_00_2} $Q_n$. Again, this definition
rests on the analogy with bound states in quantum mechanics and the
assumption of exponential localization. However, it is easy to see
that propagating modes in three-dimensional transparent periodic or
homogeneous media are all strictly non-radiating (with $\gamma_n=0$).
An example of a propagating mode in a one-dimensional periodic chain
which is strictly non-radiating, is given in Ref.~\cite{burin_04_1}.
On the other hand, radiating modes in an RPC can, in principle, be
localized. This is illustrated in Fig.~\ref{fig:2}. Here we plot the
inverse radiative quality factors $\gamma_n$ vs the corresponding
values of $M_n$ for the same set of parameters as in Fig.~\ref{fig:1}.
First, it can be seen that, while the localized modes tend to be of
higher quality, the correlation is not very strong (numerical values
of the correlation coefficient $r_c$ are indicated in each plot).
Second, there are two visibly distinct ``branches'' in
Fig.~\ref{fig:2}(a) (${\bf E}_{\rm inc}$ perpendicular to the RPC).
The top, lower quality, branch corresponds to modes with non-vanishing
dipole moments. According the criterion $M_n\ll N$, a significant
number of such modes is localized.

\section{Localization-Delocalization Transition}
\label{sec:loc-deloc}

\begin{figure}
\centerline{\psfig{file=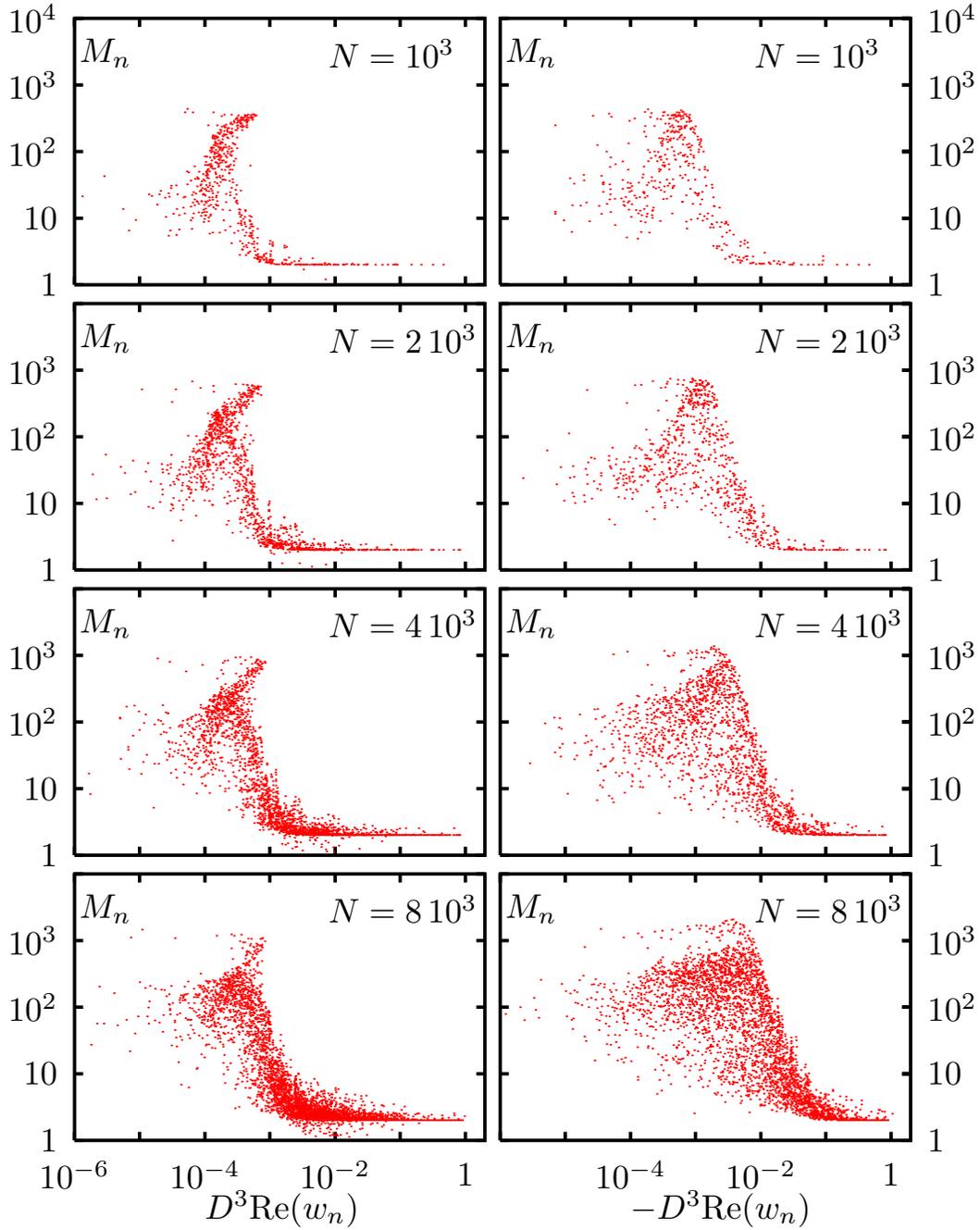,width=14cm,bbllx=165bp,bblly=360bp,bburx=450bp,bbury=742bp,clip=t}}
\caption{Participation numbers $M_n$ vs dimensionless
  spectral parameters $D^3{\rm Re}w_n$ for different densities of
  inclusions. Polarization is perpendicular to the RPC, other
  parameters same as in Figs.~\ref{fig:1} and~\ref{fig:2}.}
\label{fig:3}
\end{figure}

\begin{figure}
\centerline{\psfig{file=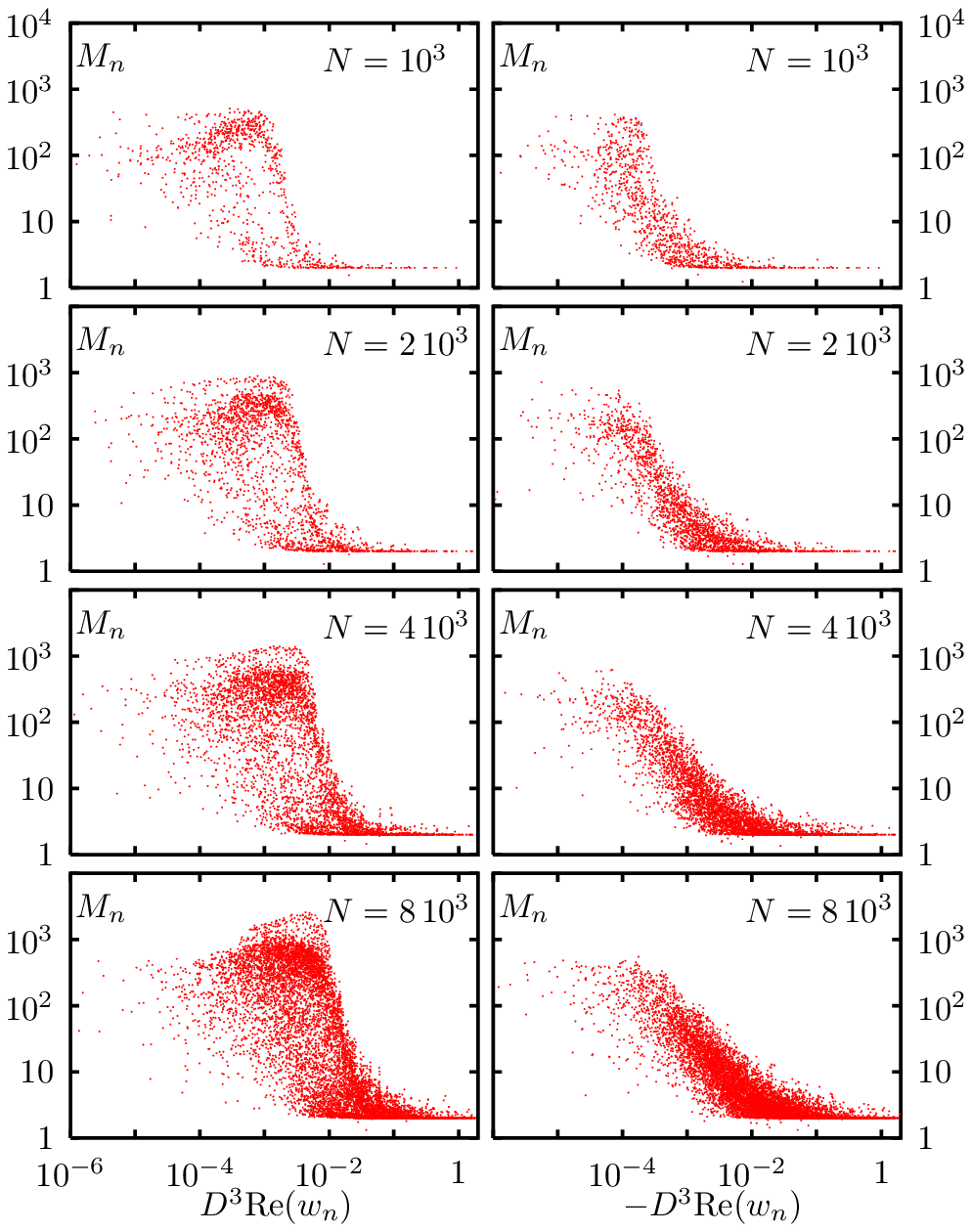,width=14cm,bbllx=165bp,bblly=360bp,bburx=450bp,bbury=742bp,clip=t}}
\caption{Same as in Fig~\ref{fig:3} but for polarization being parallel to the RPC}
\label{fig:4}
\end{figure}

So far, we have seen that some of the modes are localized on just a
few inclusions. Now we investigate if these modes actually form a
band. To this end, we plot in Figs.~\ref{fig:3} and \ref{fig:4} the
values of $M_n$ vs the appropriate spectral parameter of the theory,
which is the real part of the corresponding eigenvalue $w_n$.  To see
that ${\rm Re}w_n$ is, indeed, the spectral parameter analogous to
energy, consider the following. The $n$-th mode is resonantly excited
at an electromagnetic frequency $\omega$ such that ${\rm
  Re}[z(\omega)-w_n(\omega)]=0$ while for an isolated spherical
inclusion the resonance condition is ${\rm Re}[z(\omega)]=0$. Thus,
the real parts of the eigenvalues describe shifts of resonant
frequencies of collective excitations relative to the respective value
in the non-interacting limit. This can be illustrated with the
following simple example. Let the polarizability of a single inclusion
be given by the Lorenz-Lorentz formula with the first non-vanishing
radiative correction~\cite{draine_88_1}, namely, 

\begin{equation}
\label{z_def}
z = \frac{1}{\alpha} =
\left(\frac{2}{D} \right)^3
\frac{\epsilon+2\epsilon_h}{\epsilon-\epsilon_h} - i\frac{2k^3}{3} \ , 
\end{equation}

\noindent
where $\epsilon_h$ is the dielectric constant of the transparent host
and $\epsilon$ is the dielectric constant of the inclusions.  Further,
let $\epsilon$ be given by the Drude's formula

\begin{equation}
\label{Drude}
\epsilon(\omega) = \epsilon_0 -
\omega_p^2/\omega(\omega+i\Gamma) \ ,
\end{equation}

\noindent
where $\Gamma$ is the relaxation and $\epsilon_0$ is the intra-band
input to the dielectric function. For simplicity, let us also assume
that $\epsilon_0=\epsilon_h$ (this will not influence any conclusions
in a significant way). Then the resonance condition for the $n$-th
eigenmode takes the following form:

\begin{equation}
\label{resonance}
D^3{\rm Re}w_n = 8\left( 1 - 3\epsilon_h \omega^2/\omega_p^2 \right) \
.
\end{equation}

\noindent
Optical resonance for an isolated spherical inclusion takes place at
the Fr\"{o}hlich frequency $\omega_F = \omega_p/\sqrt{3\epsilon_h}$.
The corresponding resonance mode is characterized by ${\rm Re}w_n=0$.
Electromagnetic interaction of the inclusions results in appearance of
eigenmodes which are characterized by ${\rm Re}w_n \neq 0$.
Corresponding spectral resonances take place at frequencies different
from $\omega_F$. Using the above model for $z=1/\alpha$ and
$\epsilon$, we can estimate that the spectral shifts shown in
Figs.~\ref{fig:3},\ref{fig:4} are limited to $0.86\omega_F < \omega <
1.12\omega_F$ which corresponds to $-2<D^3{\rm Re}w_n <2$.  Note that
much larger spectral shifts can be obtained for larger densities of
inclusions. However, consideration of larger densities requires that
calculations are carried out beyond the dipole approximation.

Modes polarized perpendicular to the RPC are shown in
Fig.~\ref{fig:3}. The data for parallel polarization are shown in
Fig.~\ref{fig:4}. Analysis of Figs.~\ref{fig:3},\ref{fig:4} clearly
reveals a transition from delocalized to localized states. In
particular, all states with sufficiently large values of $D^3\vert
{\rm Re}w_n \vert$ are localized. Such states are characterized by
relatively strong interaction. In the case of low density ($N=10^3$,
$\ell \approx 32D$), most of the localized states are binary, i.e.,
involve excitation of only two inclusions. As the density of
inclusions increases, localized modes involving three, four and more
inclusions emerge. For eigenmodes polarized perpendicular to the RPC
plane and in the spectral region ${\rm Re}w_n>0$, there are also
eigenstates with $M_n=1+p$, $p\ll 1$. It can be argued that such modes
are localized on just one inclusion. Yet, ${\rm Re}w_n$ for such modes
is significantly shifted from the non-interacting limit ${\rm
  Re}w_n=0$ This result may seem to be contradictory. Indeed, if the
eigenmode amplitude is very small on all but just one inclusion, the
latter may be seen as not interacting with its environment. The
contradiction is resolved as follows. Consider a mode with eigenvalue
$w$ which is localized on the $i$-th inclusion. The corresponding
eigenvector $\vert \psi \rangle$ must then satisfy

\begin{equation}
\label{psi_eigen}
w \langle i \sigma \vert \psi\rangle = \sum_{j \neq
  i} \sum_{\tau} \langle i \sigma \vert W \vert j \tau
  \rangle \langle j\tau \vert \psi \rangle \ .
\end{equation}

\noindent
Here $\sigma, \tau$ label the Cartesian components of vectors and the
index $n$ that labels eigenmodes is omitted; we thus focus attention
only on the selected eigenstate. The eigenmode is localized on the
$i$-th inclusion if $m({\bf r}_i) = 1-p$ where $p \ll 1$. Therefore
$m({\bf r}_i) \sim 1$ and, since the weights satisfy the sum rule
(\ref{m_n_norm}), we also have $m({\bf r}_j) \sim p/N$ for $j \neq i$.
We now recall that the weights are quadratic in eigenvector
components.  Consequently, $\langle j \tau \vert \psi \rangle \sim
\sqrt{p/N}$ for $j \neq i$ while $\langle i \sigma \vert \psi \rangle
\sim 1$. It then follows from (\ref{psi_eigen}) that

\begin{equation}
\label{single_scaling}
w \sim N \langle W \rangle \sqrt{p/N} =  \langle W \rangle \sqrt{pN}\
,
\end{equation}

\noindent
where $\langle W \rangle$ is the appropriate average of the
interaction operator in the right-hand side of (\ref{psi_eigen}) and
the above relation is accurate only to the order of magnitude. We thus
see that, even if $p$ is arbitrarily small, the spectral shift may not
be small in a sufficiently large sample (large $N$). This result is
not specific to the RPC's but is valid in two and three-dimensional
disordered media as well. Of course, the value of $\langle W \rangle$
will depend on the dimensionality of the sample and on the density of
inclusions.  We note that $\langle W \rangle$ is the same as the
factor $Q$ introduced by Berry and Percival within the mean-field
approximation~\cite{berry_86_1}.

The phenomenon of spectrally shifted eigenstates which are localized
on just one inclusion is explained by constructive interference and
has no counterpart for electrons in disordered solids. This is because
the analogy between the spectral parameter ${\rm Re} w_n$ and energy
is not complete.  Indeed, in the case of the classic Anderson model,
an electronic state can be localized at an anomalously deep local
potential. Such state is not influenced in any way by values of
potentials at neighboring sites since the electron is exponentially
localized inside the potential well. But the localization phenomenon
discussed here is, essentially, collective and depends on the
particular realization of the random sample as a whole.  Likewise, the
binary states seen in Figs.~\ref{fig:3},\ref{fig:4} are not
necessarily binary states of two closely situated inclusions which
interact with the rest of the sample very weakly (although such states
are also possible; see Ref.~\cite{markel_92_1} for properties of
isolated dimer states).  This is evident already from the data shown
in Fig.~\ref{fig:1}. Here some of the binary states (with $M_n \approx
2$) are characterized by large gyration radii $\xi_n$ and, therefore,
are not localized on two closely placed inclusions.

It can also be seen that there is a spectral region (which depends on
the density of inclusions and on polarization of the eigenmodes) where
localized and delocalized modes co-exists simultaneously. This
spectral region is especially well pronounced in the case of larger
densities and for polarization of eigenmodes parallel to the plane of
the RPC.  Co-existence of localized and delocalized modes with very
close spectral parameters has been previously demonstrated for random
three-dimensional fractal aggregates and was referred to as
inhomogeneous localization~\cite{stockman_96_1}. Inhomogeneous
localization was also observed in the case of
RPCs~\cite{stockman_01_1}. However, the findings of
Refs.~\cite{stockman_01_1,stockman_96_1} were based on the
quasistatic approximation and on the definition (\ref{xi_def}) of the
localization length. It was concluded that localization is
inhomogeneous in the whole spectral range. Here we show that
inhomogeneous localization takes place only in a transitional energy
band. The width of this transitional band as a function of system size
and density of inclusions needs to be further investigated,
especially, without the use of dipole approximation.

\section{Coupling of Localized Models to the Far Field}
\label{sec:coupling}

\begin{figure}
\centerline{\psfig{file=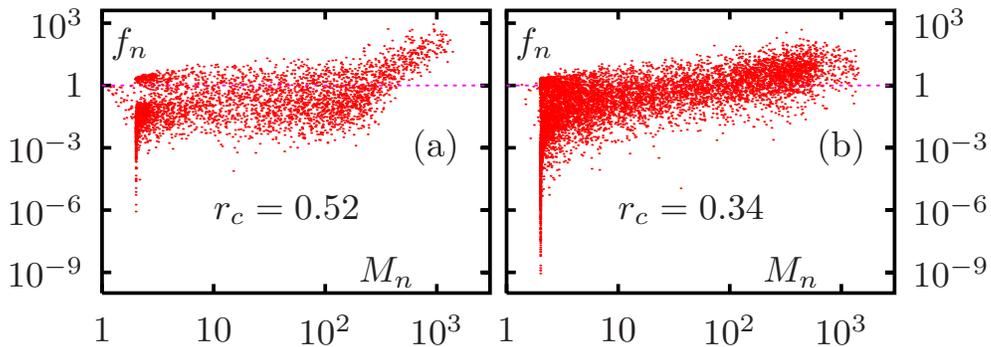,width=14cm,bbllx=165bp,bblly=620bp,bburx=450bp,bbury=742bp,clip=t}}
\caption{Coupling constants $f_n$ vs participation numbers
  $M_n$. Coupling constants are computed for an external plane wave
  ${\bf k}_{\rm inc}=k\hat{\bf x}$ and ${\bf E}_{\rm inc}=E_0\hat{\bf
    z}$ (a) and for ${\bf k}_{\rm inc}=k\hat{\bf z}$ and ${\bf E}_{\rm
    inc}=E_0(\hat{\bf x} + \hat{\bf y})/\sqrt{2}$ (b).  Other
  parameters same as in Fig.~\ref{fig:1}. Numerical value of the
  correlation coefficient $r_c$ between $M_n$ and $f_n$ is indicated
  in each plot.}
\label{fig:5}
\end{figure}

We now consider the possibility of coupling of localized modes to the
external field. It has been previously argued that, in the quasistatic
limit, strongly localized modes can not be effectively coupled to
plane waves. Therefore, such modes were referred to as
dark~\cite{stockman_01_1}. However, the dark modes become coupled to
external plane waves if one considers first non-vanishing corrections
in $k$, i.e., goes beyond the quasistatic limit. For example, in
Ref.~\cite{markel_92_1} it was shown that the fully antisymmetrical
mode of two oscillating dipoles (with zero total dipole moment) can be
effectively coupled to an external plane wave in the limit $kL
\rightarrow 0$. This coupling is explained by a small phase shift of
the incident wave and the high quality factor of the mode. Indeed, it
can be seen from (\ref{d_spectral}) that under the exact resonance
condition ${\rm Re}(z-w_n)=0$, the excited dipole moments become
proportional to $f_n^{\rm (eff)}=Q_n f_n$, where $Q_n=1/\gamma_n$.
Thus, even in samples that are small compared to the wavelength,
high-quality modes with zero or vanishing dipole moment can be
effectively excited under the resonance condition. When the sample
size is not small compared to the wavelength, even the strict
resonance condition is not required for effective coupling.  This is
illustrated in Fig.~\ref{fig:5}. Here we plot the coupling constants
$f_n$ vs $M_n$ for the same set of parameters as in Figs.~\ref{fig:1}
and \ref{fig:2}. Since the coupling constants are normalized by the
condition $\sum_n f_n = N$, a mode is coupled {\em effectively} if
$f_n \sim 1$. The modes with $f_n \ll 1$ are coupled {\em weakly} and
the modes with $f_n\sim N$ are coupled {\em strongly}. Correlation
between $M_n$ and $f_n$ appears to be quite weak (see the figure for
numerical values of the correlation coefficient $r_c$). Most
importantly, it can be seen that a considerable fraction of localized
modes is {\em effectively} coupled to the external wave, although only
delocalized modes can be coupled {\em strongly}.

\begin{figure}
\centerline{\psfig{file=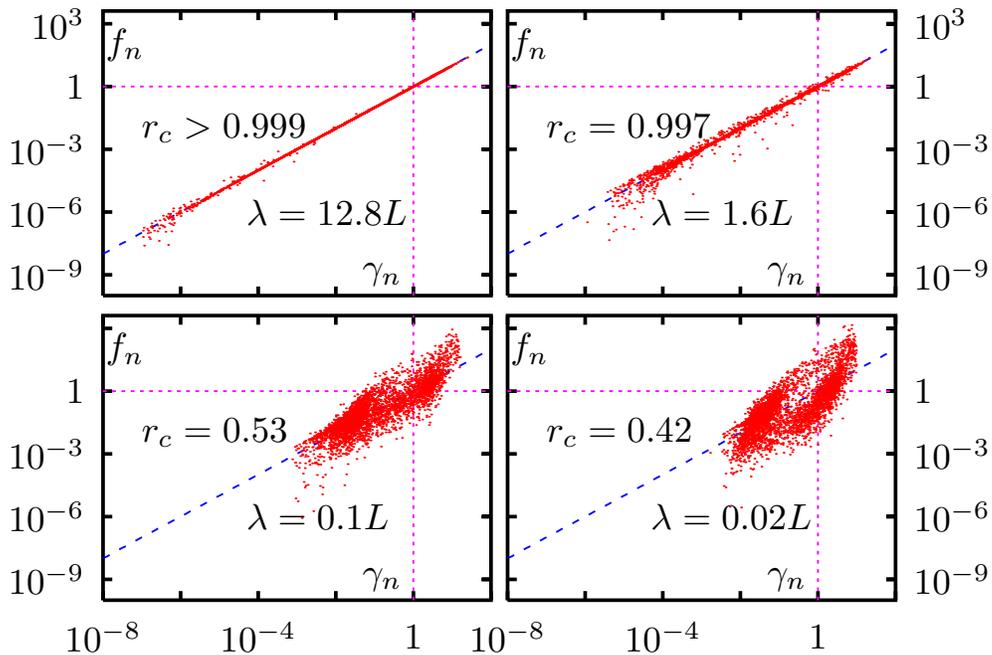,width=14cm,bbllx=165bp,bblly=543bp,bburx=450bp,bbury=742bp,clip=t}}
\caption{Coupling constants $f_n$ vs inverse radiative
  quality factors $\gamma_n$ for all modes polarized perpendicular to
  the RPC with $N=4000$ inclusions and different ratios $\lambda/L$,
  as indicated. In all cases, $L=1000D$. The dashed blue line
  corresponds to the quasistatic result $f_n=\gamma_n$. Numerical
  values of the correlation coefficient $r_c$ between $\gamma_n$ and
  $f_n$ are indicated.}
\label{fig:6}
\end{figure}

\begin{figure}
\centerline{\psfig{file=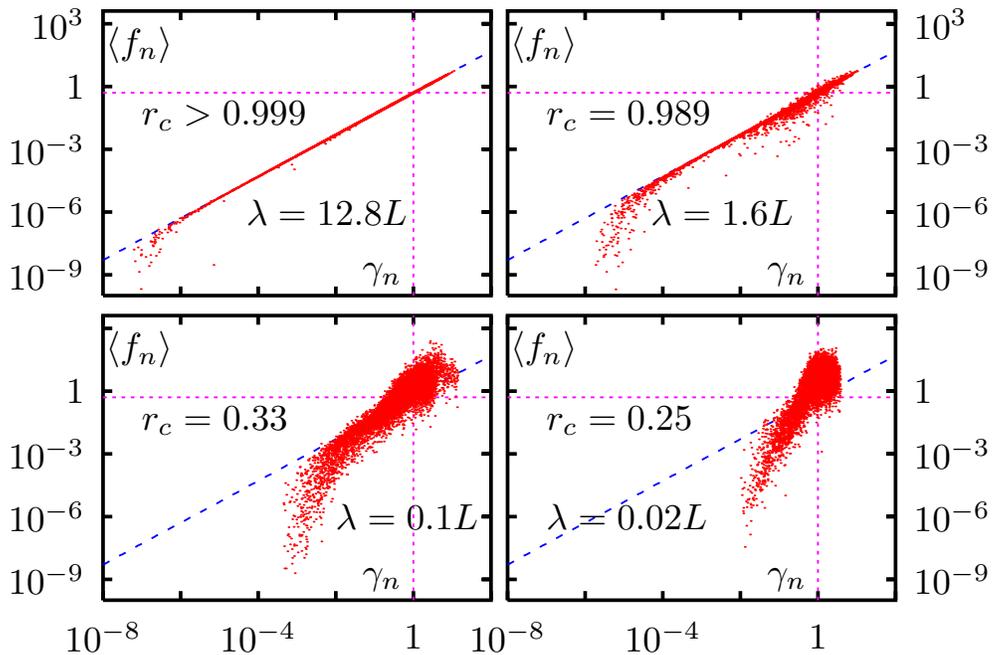,width=14cm,bbllx=165bp,bblly=543bp,bburx=450bp,bbury=742bp,clip=t}}
\caption{Same as in Fig.~\ref{fig:6} but for modes polarized in plane
  of the RPC and the coupling coefficients $f_n$ are averaged over the
  two orthogonal in-plane polarizations of the incident wave. The
  dashed blue line corresponds to the quasistatic result $\langle f_n
  \rangle=\gamma_n/2$ and $\langle \ldots \rangle$ denotes
  polarization averaging. Numerical values of the correlation
  coefficient $r_c$ between $\gamma_n$ and $\langle f_n \rangle$ are
  indicated.}
\label{fig:7}
\end{figure}

Perhaps, the most counter-intuitive fact about the polar eigenmodes
that can be understood only beyond the quasistatics is that the
inverse radiative quality factor $\gamma_n$ and the coupling constant
$f_n$ are not necessarily proportional to each other.  In
Figs.~\ref{fig:6},\ref{fig:7} we plot $f_n$ vs $\gamma_n$ for
different ratios $\lambda/L$ and for mode polarization perpendicular
to the RPC (Fig.~\ref{fig:6}) and parallel to the RPC
(Fig.~\ref{fig:7}). First, consider the case of orthogonal
polarization. In the quasistatic limit, it can be
shown~\cite{markel_95_1} that (for this particular polarization of the
eigenmodes) $f_n=\gamma_n$.  This proportionality is clearly visible
in the case $\lambda/L=12.8$ and the correlation coefficient $r_c$
between $\gamma_n$ and $f_n$ exceeds $0.999$. However, at smaller
values of $\lambda/L$, there is no strict proportionality.  Thus, for
example, in the case $\lambda/L=0.02$, the correlation coefficient is
rather small ($r_c=0.42$). It can also be seen that the modes with
$\gamma_n\approx 1$ have coupling constants which differ by four
orders of magnitude and can be either weakly or strongly coupled to
the far field.  Likewise, modes that are effectively coupled to the
far field ($f_n\approx 1$) can be either weakly radiating
($\gamma_n\ll 1$) or strongly radiating ($\gamma_n \gg 1$). 

Now consider eigenmodes which are polarized in the RPC plane
(Fig.~\ref{fig:7}). There are two linearly-independent in-plane
polarizations of the incident wave, ${\bf E}_{\rm inc}$. While the
radiative factors are independent of ${\bf E}_{\rm inc}$, this is not
so for $f_n$'s. For any given direction of ${\bf E}_{\rm inc}$, there
is no strict proportionality between $\gamma_n$ and $f_n$ even in the
quasistatic limit. This lack of proportionality is a consequence of
polarization effects. For example, there might be modes which are
strongly coupled to incident waves polarized along the $x$-axis but
not coupled to waves polarized along the $y$-axis. However, the
polarization effects can be suppressed by taking the average of $f_n$
over two linearly-independent incident polarizations. We denote such
average as $\langle f_n \rangle$ and it can be shown that in the
quasistatic limit $\langle f_n \rangle = \gamma_n/2$. This is indeed
confirmed by the data shown in Fig.~\ref{fig:7}. Qualitatively, the
data in Fig.~\ref{fig:7} are similar to those shown in
Fig.~\ref{fig:6}, with even smaller correlation factors.

The finding that weakly radiating modes can be effectively coupled to
propagating waves is counter-intuitive and even may seem to contradict
conservation of energy. Indeed, consider excitation of a mode which is
effectively coupled to the far field but is weakly radiating by an
electromagnetic wave that is ``turned on'' at an initial moment of
time $t=0$. Since the mode is weakly radiating, it would not
contribute significantly to the scattered field. Thus the incident
wave would seemingly pass through the sample without noticeable
scattering or absorption. However, at a sufficiently large time $t=T$,
a steady state would be reached, in which a finite electromagnetic
energy would be transferred to plasmonic oscillations in the mode.
Since the incident wave was not scattered or absorbed, this
contradicts energy conservation. In fact, the contradiction is
resolved by noticing that a given mode is weekly radiating only at a
fixed electromagnetic frequency $\omega$. (Beyond the quasistatic
limit, both the eigenvectors $\vert \psi_n\rangle$ and the eigenvalues
$w_n$ are functions of $\omega$.) But the transition process described
above necessarily involves incident waves of different frequencies,
not all of which would pass through the sample without scattering.
This would result in non-zero extinction of the incident power.

\section{Summary and Discussion}
\label{sec:summary}

Localization of polar eigenmodes in random planar composites (RPCs)
has been studied theoretically and numerically without the quasistatic
approximation. It was demonstrated that the localization criteria
based on exponential confinement (analogy with electrons in solids)
can not be applied to polar excitations in disordered composites and
in RPCs in particular. This is because localization of polar
eigenmodes is algebraic rather than exponential. Still, eigenstates
with algebraically decaying tails can be square-integrable and
discrete, and therefore, localized in the Anderson sense.

Note that localized eigenstates whose tails decay according to power
law have been also discovered for Hamiltonians which can be
represented by so-called random banded matrices with algebraically
decaying bands~\cite{mirlin_96_1}. Elements of such matrices $a_{ij}$
decay as $\vert i - j \vert^{-\alpha}$. Here $\vert i - j \vert$ can
be viewed as a liner distance in a one-dimensional system.  It was
found that for $\alpha<1$ all eigenstates are delocalized while for
$\alpha>1$ all eigenstates are algebraically localized. In the
critical case $\alpha=1$, the structure of eigenstates was found to be
multifractal.  These findings were in agreement with the results
obtained earlier for random mechanical oscillators coupled by
quasistatic dipole-dipole interaction in
Refs.~\cite{levitov_89_1,levitov_90_1}. In these references, the
long-range interaction was accounted for perturbatively, but arbitrary
dimensionality of space $d$ was considered, with the conclusion that
all eigenstates are delocalized when $\alpha \leq d$. We note that
there are several important physical differences between the model
considered here and in
Refs.~\cite{levitov_89_1,levitov_90_1,mirlin_96_1}. For example,
in~\cite{levitov_90_1} two mechanical oscillators are in resonance
irrespectively of the distance between them if their frequencies
exactly coincide. In the model discussed here, resonances are not
mechanical but electromagnetic, with resonance frequency strongly
depending on the geometrical arrangement of
dipoles~\footnote[2]{Strictly speaking, the appropriate spectral
  parameter for the problem of collective electromagnetic excitations
  discussed in this paper is $z(\omega)$ given by Eq.~(\ref{z_def})
  rather than the frequency $\omega$ itself, with resonances taking
  place at frequencies satisfying one of the equations ${\rm
    Re}[z(\omega)- w_n]=0$.}. Instead of solving the problem of weakly
coupled oscillators of random frequencies, we solve the problem of
electromagnetically coupled, identical polarizable particles.
Further, a random banded matrix of size $N$ has $N^2$ mathematically
independent elements while the interaction operator $W$ of size $N$
(studied in this paper) depends on only $N$ mathematically independent
variables ${\bf r}_i$ and is not banded.  However, the mathematical
reason for algebraic rather than exponential localization appears to
be similar in both models: the algebraic spatial decay of interaction.
Theory of localization for systems whose Hamiltonian can be
represented by power-law band random matrices was further developed in
Refs.~\cite{mirlin_00_1,varga_02_1,cuevas_02_1,cuevas_03_1}.  It is
interesting to note that electromagnetic interaction in the radiation
zone decays as $r^{-1}\exp(ikr)$ and $d=2$ for the RPCs. Thus, the
rate of algebraic decay corresponds to the regime $\alpha < d$.
However, the interaction is modulated by the exponential phase factor
$\exp(ikr)$ which makes a direct comparison of results problematic.

Since localized states in an RPC have algebraically decaying tails,
the previously used localization criterion based on the gyration
radius of the mode is inapplicable. Consequently, an alternative
approach based on the participation number has been used in this
paper. It was shown that all electromagnetic states in the RPC whose
resonance frequencies are shifted from those of non-interacting
inclusions by a value larger than certain threshold are localized. The
band of localized states shown in Figs.~\ref{fig:3} and~\ref{fig:4}
for sufficiently large values of $\vert {\rm Re}(w_n) \vert$ can be
mapped to an interval of electromagnetic frequencies if the material
properties of the inclusions and the host medium are specified. It
should be noted that much stronger spectral shifts will be observed at
higher concentrations of inclusions. Consideration of the high-density
limit will require solving the electromagnetic problem without the
dipole approximation. When applied to large random systems, this is a
very computationally demanding procedure, solution to which, at least
at the time being, appears to be not feasible. The author expects that
this will not influence the localization properties of the eigenmodes.

Finally, possibility of coupling of localized modes to the far field
has been studied. It was shown that, contrary to the previous belief,
localized modes in the RPCs can be effectively coupled to the far
field.

\section*{References}
\bibliography{abbrev,master,local}

\end{document}